\documentclass[5p,twocolumn]{elsarticle}
\usepackage{graphicx,latexsym}
\usepackage{dcolumn}
\usepackage{amssymb,amsmath,bm}
\usepackage{subfigure}
\usepackage{braket}
\usepackage{siunitx}
\usepackage{array}
\usepackage{float}
\usepackage[nopar]{lipsum}
\usepackage{caption}
\usepackage{multirow}
\usepackage{bigstrut}
\usepackage[super]{nth}

\newcommand{\angstrom}{\text{\normalfont\AA}}
\usepackage{hyperref}
\hypersetup{
    pdfnewwindow=true,       % links in new window
    colorlinks=true,         % false: boxed links; true: colored links
    linkcolor=blue,          % color of internal links
    citecolor=blue,          % color of links to bibliography
    filecolor=magenta,       % color of file links
    urlcolor=black           % color of external links
}

\usepackage[normalem]{ulem}

\def\fig#1{Fig.\ \ref{#1}}
\def\tab#1{Tab.\ \ref{#1}}

\journal{}

\begin{document}

\begin{frontmatter}

%-----------------------------------------------------------------

%\title{Spin-polarized DFT study of tunable electronic, thermal, optical and magnetic properties of monolayer Ga$_2$Si$_6$} 

\title{DFT study of tunable electronic, magnetic, thermal, and optical properties of a Ga$_2$Si$_6$ monolayer}

\author[a1,a2]{Nzar Rauf Abdullah}
\ead{nzar.r.abdullah@gmail.com}
\address[a1]{Division of Computational Nanoscience, Physics Department, College of Science, 
             University of Sulaimani, Sulaimani 46001, Kurdistan Region, Iraq}
\address[a2]{Computer Engineering Department, College of Engineering, Komar University of Science and Technology, Sulaimani 46001, Kurdistan Region, Iraq}

\author[a3]{Botan Jawdat Abdullah}
\address[a3]{Physics Department, College of Science, Salahaddin University-Erbil, Erbil 44001, Kurdistan Region, Iraq
}

\author[a5]{Vidar Gudmundsson}
\address[a5]{Science Institute, University of Iceland, Dunhaga 3, IS-107 Reykjavik, Iceland}

%----------------------------------------------------------------

\begin{abstract}

The electrical, magnetic, thermal and optical characteristics of Gallium (Ga) doped silicene are investigated using density functional theory (DFT). The effect of doping is studied by tuning dopant concentrations as well as examining varied doping distances, and atomic dopant interactions for the same substitutional doping concentration. 
The results indicate that the Ga atoms alter the band structure and the band gap in the silicene monolayer at various concentrations, which can be referred back to to the repulsive interaction of Ga-Ga atoms. 
The band gap is determined by the interaction strength of the Ga-Ga atoms, the Coulomb repulsive force, and it does not always widen as doping concentration increases. In addition, our spin-polarized DFT calculations show that these monolayers behave like nonmagnetic semiconductors, exhibiting symmetric spin-up and spin-down channels. 
The repulsive interaction between the Ga atoms causes a symmetry breaking of the monolayers.
As a consequence, a Ga dopant can open the band gap, leading to better thermoelectric properties such as the Seebeck coefficient and the figure of merit, as well as an increase in the optical response. As a result of our estimates, Ga doped silicene monolayers could be advantageous in thermoelectric and optoelectronic devices.

\end{abstract}


\end{frontmatter}

\section{Introduction} 

Two-dimensional (2D) nanomaterials, which are made up of small systems with a thickness of at least one atomic layer, have emerged as an extraordinary class of materials. These 2D materials have a high surface-area-to-volume ratio as compared to bulk materials. A very thin sheet’s magnetic, electrical, optical, mechanical, and catalytic properties may be fine-tuned by carefully controlling the size, shape, synthesis conditions, and functionalization \cite{D0MA00807A, Kim_2018, ABDULLAH2021413273, doi:10.1021/nn400280c}.
Since the successful fabrication of graphene, the first two-dimensional (2D) material, scientists have been working on it continually, both theoretically and experimentally. Other graphene-related compounds are apparently still being explored in order to broaden the scope of possible device applications \cite{Yu2017, C7TA00863E}. Silicene is a silicon monolayer with a low-buckled honeycomb and a Dirac cone in the band structure that is the silicon counterpart of graphene. Due to its amazing properties for a wide range of applications and relative ease of integration with today's current silicon-based electronic devices, 2D silicene monolayer is one of the most fascinating two-dimensional materials \cite{OUGHADDOU201546, Tokmachev2018, PhysRevB.76.075131, doi:10.1063/1.4944631}.
Monolayer silicene has been successfully synthesized so far using a variety of experimental techniques. The bottom-up technique has been the most common, including epitaxial growth on a substrate by deposition onto a supporting template such as monolayer silicence on Ag (111) substrate \cite{PhysRevLett.108.155501, Lin_2012, Jamgotchian_2012, doi:10.1021/nl301047g, ARAFUNE2013297}, also fabricated on zirconium diboride thin films \cite{PhysRevLett.108.245501}, Ir (111) \cite{doi:10.1021/nl304347w},  ZrB$_2$ \cite{Aizawa_2015}, MoS$_2$\cite{Jiajie_Zhu_2015, https://doi.org/10.1002/adma.201304783} and SiC \cite{doi:10.1021/jp311836m} substrate. Furthermore, Tao et al.\ \cite{Tao2015} recently confirmed the development of monolayer silicene field-effect transistors based on Ag(111). As a result, the substrate's role in silicene's properties is important for practical uses of silicene.

Unlike graphene, which is flat and stable, silicene is somewhat sticky and hence unstable due to its puckered or crinkled structure \cite{Takahashi2017}. Despite the fact that their crystal structures are identical, silicene creates sp$^3$ bonds, whereas graphene forms sp$^2$ bonds. Their applications will be limited because of the lack of a band gap due to its buckled structure \cite{C8CS00338F}. A key success for enhancing silicene-targeted applications is an increase in its free-carrier density or the opening of its band gap. Despite the fact that external electric, magnetic, and mechanical forces all have a significant impact on the electrical properties of silicene, doping more significantly affects the properties of silicene \cite{doi:10.1021/ar400180e, C5CP04841A, Nguyen2019, doi:10.1021/nn504451t}. 

Several theoretical studies of silicene have recently indicated that it is very desirable for enhancing device performance and remains a challenge to further studies via doping. For example, in the framework of DFT within GGA, transition metal doped silicene Vanadium (V) can produce a stable quantum anomalous Hall effect and provide a platform for electrically controlled topological states \cite{Zhang2013}, the energy bands at the Dirac point in Ni-doped silicene undergo a transformation from linear to parabolic dispersion, which is further stabilized \cite{C3CP54655A}.

Furthermore, a new extra electron energy loss spectra peak is found in perpendicular polarization for the optical characteristics of free standing silicene with different concentrations of Al, P, and Al-P codoping \cite{C4RA07976K}, apart from the Fermi level, which varies on the concentration of the atom doped and preserves the linear energy dispersion, the Al, B, and P interact significantly with the Si atoms of the dopant atoms in silicene \cite{HERNANDEZCOCOLETZI2018242} and controlling the doping locations in the structures changes dramatically the electronic characteristics of C doped silicene monolayers, resulting in a Mott-Anderson transition \cite{doi:10.1021/acs.nanolett.0c01775}.

In addition, transition metals (TM) such as Ti, V, Mn, Fe, and Co lower the buckling degree of the silicene structure, and significant orbital hybridization between TM atoms and silicene creates a high magnetic moment. In the absence of spin-polarization, certain TM-silicenes exhibits a high Seeback coefficient \cite{ABDULLAH2021114644}, light rare-earth (RE) metals doped in silicene, such as La, Ce, Pr, Nd, Pm, and Sm, can create significant magnetic moments, and their magnetic characteristics are largely produced by RE atoms in their f states, with the exception of Pm and Sm doped silicene \cite{LI2020106712}, and under certain adsorption configurations, metals such as Na, Mg, and Al in doped silicene enhanced the buckling degree, the free-carrier densities, changed the Dirac cone structure, the spin-degenerate electron energy spectrum across Fermi energy, and the ferromagnetic properties \cite{doi:10.1021/acsomega.0c00905}.

In this work, monolayer Silicene doped with Ga forming a Ga$_2$Si$_6$ monolayer is investigated with different concentrations and atomic configurations of Ga atoms and the interaction between Ga atoms. The obtained results are analyzed using the spin-dependent and independent DFT forms. The band gap is opened when the distance between the doping atoms is large or the repulsive between the Ga atom is weak, causing silicene to become a semiconductor, and it tends to zero when they are near one other due to the strong repulsive force. Thus, the electrical, thermal, optical, and magnetic characteristics of a silicene monolayer are enhanced by the repulsive interactions between the dopant atoms, which has not previously been reported. In this research we discover that Ga doping of silicene has a novel effect on the physical characteristics of silicene monolayers, which is quite attractive for future nano-optoelectronics applications.

\section{Model and Method}

All computations of electrical and optical properties 
are carried out using DFT, as implemented in the Quantum Espresso (QE) code \cite{Giannozzi_2009, giannozzi2017advanced}. Using the generalized gradient approximation (GGA) in the Perdew-Burke-Ernzernhof (PBE) functionals, the electronic band structure and optical properties are studied. The (XCrySDen) \cite{KOKALJ1999176} and VESTA \cite{momma2011vesta} are used in this work to visualize the crystalline and molecular structure of the Ga doped silicene monolayers.
We use a $18\times18\times1$ Monkhorst-Pack K-point mesh with a $1088$~eV energy cutoff for fully relaxed monolayers \cite{ABDULLAH2021106981}. The atomic relaxation is performed until all the forces on the atoms are less than $10^{-5}$ eV/$\angstrom$. The density of states (DOS) calculations are done using a $100 \times 100 \times 1$ K-mesh grid points. Finally, the thermoelectric properties are determined using the BoltzTraP software program \cite{Madsen2006, ABDULLAH2021106073}. The BoltzTraP code has a QE interface and functions with a mesh of band energies.

\section{Results and discussion}
 
The electrical, magnetic, thermal, and optical properties of Ga doped silicene are computed and analyzed in this section. Various concentrations and configurations of silicene at different doping levels are computed in a $2\times2$ supercell.

\subsection{Electronic properties} 

In general, a $2\times2$ supercell has three atomic positions: ortho, meta, and para positions \cite{RASHID2019102625, ABDULLAH2021110095}.  If one Ga atom is doped in silicene identified as GaSi$_7$, the concentration of the Ga dopant atom is $12.5\%$. The position of Ga atom in all the three positions of the hexagon of silicene will give similar physical properties.  If two Ga atoms are doped in the silicene identified as Ga$_2$Si$_6$, the concentration of the Ga dopant atoms is $25\%$. In this case three distinct substitutional doping possibilities are considered. These three doping possibilities are expected to have different physical properties. 
\begin{figure}[htb]
	\centering
	\includegraphics[width=0.5\textwidth]{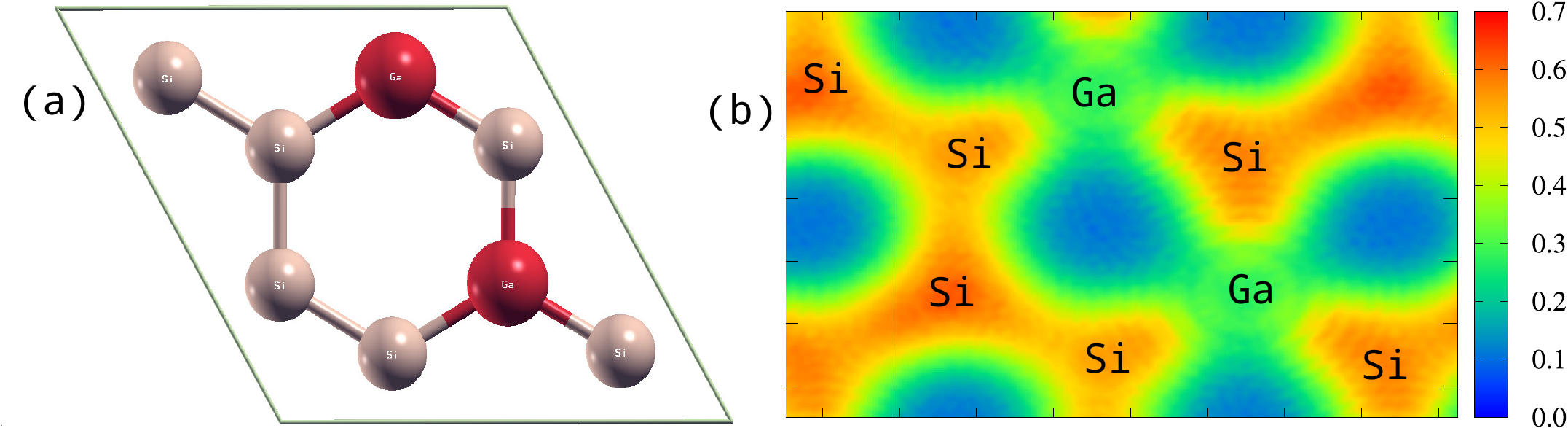}
	\caption{(a) Atomic configuration, and (b) electron localization function of Ga$_2$Si$_6$-2 monolayer. Red and brown color balls are Ga and Si atoms, respectively.}
	\label{fig01}
\end{figure}
The two Ga atoms can be doped at para-ortho, para-meta, and para-para positions. These three different atomic configurations are identified as Ga$_2$Si$_6$-1, Ga$_2$Si$_6$-2, and Ga$_2$Si$_6$-3 for para-ortho, para-meta, and para-para positions, respectively. 
For instant, the atomic configuration (a) and the electronic localization function (b) of one of these structures, Ga$_2$Si$_6$-2 is shown in \fig{fig01}, where the two Ga atoms are doped at the para-meta
positions.

It seems that the electron charge transfers from the Ga atoms to the Si atoms in all monolayers under investigation here, which is expected as the electronegativity of an Si atom, $1.9$, is higher than that of a Ga atom, $1.81$.

The energetic stability of the Ga doped silicene monolayers is gauged by a calculation of their formation energy \cite{ABDULLAH2020114556}. The DFT calculations indicate that the three Ga$_2$Si$_6$ monolayers can be arranged from the higher to lower formation energy as follow: Ga$_2$Si$_6$-1 $<$ Ga$_2$Si$_6$-3 $<$ Ga$_2$Si$_6$-2.
It has been shown that the lower the formation energy, the more energetically stable structure is found. As a result, the Ga$_2$Si$_6$-1 monolayer is the most stable structure. 

The Ga atom doped in silicene will modify the electronic structure such as the lattice constant, the bond lengths, and the band gap. \tab{table_one} presents the physical parameters of the four configurations, that have been considered here. We determine, that the lattice constants of GaSi$_7$, Ga$_2$Si$_6$-1, Ga$_2$Si$_6$-2, and Ga$_2$Si$_6$-3 are longer than that of pure silicene. The lattice constant of pure silicene is $3.86 \angstrom$ \cite{abdullah2021interaction}. Comparing to the lattice constant of pure silicene, the lattice constant of Ga-doped silicene is larger (see \tab{table_one}) indicating a supercell expansion of these Ga doped silicene monolayers.
In comparison to them, the Ga$_2$Si$_6$-3 structure has a larger lattice constant between atoms than the others. 
\begin{table}[h]
	\centering
	\begin{center}
		\caption{\label{table_one} Lattice constant, $a$, Si-Si, Ga-Si, and Ga-Ga bond length, and band gap for all Ga doped silicene structures. The unit of all bond length is $\angstrom$.}
		\begin{tabular}{|l|l|l|l|l|l|l|}\hline
			Structure	      & a       & Si-Si & Ga-Si  & Ga-Ga   & band gap    \\ \hline
			GaSi$_7$	      & 3.95    & 2.31  &  2.31  &  -      &  0.28 eV  \\
			Ga$_2$Si$_6$-1	  & 4.03    & 2.28  &  2.35  &  2.43   &   0.2 meV \\ 
			Ga$_2$Si$_6$-2	  & 4.04    & 2.31  &  2.34  &  -      &  0.12 eV  \\
			Ga$_2$Si$_6$-3	  & 4.05    & 2.25  &  2.38  &  -      &  0.6 eV  \\   \hline
		\end{tabular}
	\end{center}
\end{table}
The supercell expansion arises from the Ga-Ga interaction in the silicene structure. 
To determine the interaction type between the Ga-Ga atoms \cite{ABDULLAH2020126350}, our DFT calculations show that the interaction energy between the Ga atoms is repulsive as the interaction energy is positive. The repulsive interaction is directly proportional to the distance between Ga atoms. The repulsive interaction energy is $2.18$, $1.6$, and $0.9$~eV for Ga$_2$Si$_6$-1, Ga$_2$Si$_6$-2, and Ga$_2$Si$_6$-3, respectively. The modifications in the lattice constant and the bond length influence the symmetry of the structure and thus the band structures. 

\begin{figure}[htb]
 	\centering
 	\includegraphics[width=0.45\textwidth]{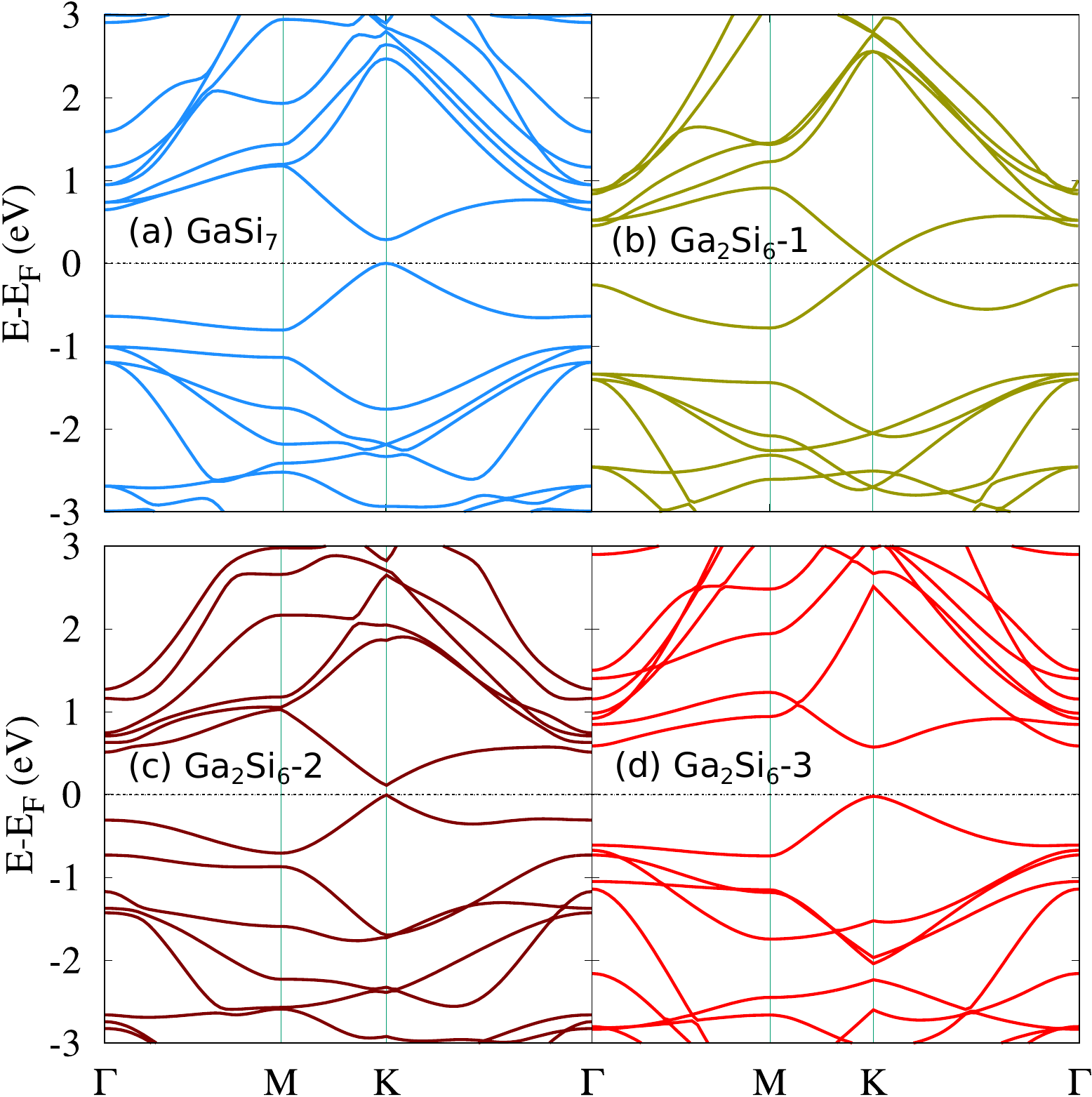}
 	\caption{Band structure for optimized structures of GaSi$_7$ (a), Ga$_2$Si$_6$-1 (b), and  Ga$_2$Si$_6$-2 (c), and  Ga$_2$Si$_6$-3 (d).
 	The energies are with respect to the Fermi level, and the Fermi energy is set to zero.}
 	\label{fig02}
 \end{figure}

It has been shown that the 2D pure silicene monolayer has no band gap, similar to graphene, and the valence and conduction bands meet at the Fermi level \cite{PhysRevB.50.14916, ABDULLAH2020126807}. The band structure of silicene doped with Ga atoms is determined using DFT methods. 
The electronic band structure of the GaSi$_7$, Ga$_2$Si$_6$-1, Ga$_2$Si$_6$-2, and Ga$_2$Si$_6$-3 systems is shown in \fig{fig02}. The doping was initially investigated at low impurity concentrations. When one Si atom is replaced with one Ga atom in a supercell, the impurity concentration rises to $12.5\%$. Our band gap of GaSi$_7$ is $0.28$~eV, as shown in \fig{fig02}(a), which is in good agreement with previous theoretical result of $0.275$~eV \cite{refId0}. 
The Ga atom alters the band gap, causing silicene to transform from semimetal to semiconductor. Due to the presence of a finite bandgap around Fermi energy, silicene doped with Ti and V is a semiconductor, whereas Mn, Fe, and Co doped silicene is a metal because the Fermi energy penetrates the valence or the conduction bands.

It has been shown that the band gap of most monolayers is increased with the dopant concentration ratio \cite{C4NR00028E, C6RA04782C, ABDULLAH2020100740}. This status is not true if one considers the interaction between the dopant atoms.
We investigated Ga doping at $25\%$ impurity concentration with various Ga-Ga separations by substituting two Si atoms with two Ga atoms. There are three possible doped silicene monolayers in supercell with two Ga atoms in various positions, yielding three distinct lattice constants, and interaction energy as is seen in \tab{table_one}. Our findings show that the band gap is determined by the repulsive interaction between the Ga doped atoms, rather than by the amount of impurities. The band gap opens as the distance between the Ga-Ga atoms is slightly increased, while the band gap narrows as the repelling force between them increased.
% 
% \lipsum[0]

The band structure is significantly tuned, and a band gap is opened based on Ga atom configurations for the GaSi$_7$, Ga$_2$Si$_6$-1, Ga$_2$Si$_6$-2, and Ga$_2$Si$_6$-3 structures.
\lipsum[0]
\begin{figure*}[htb]
	\centering
	\includegraphics[width=0.8\textwidth]{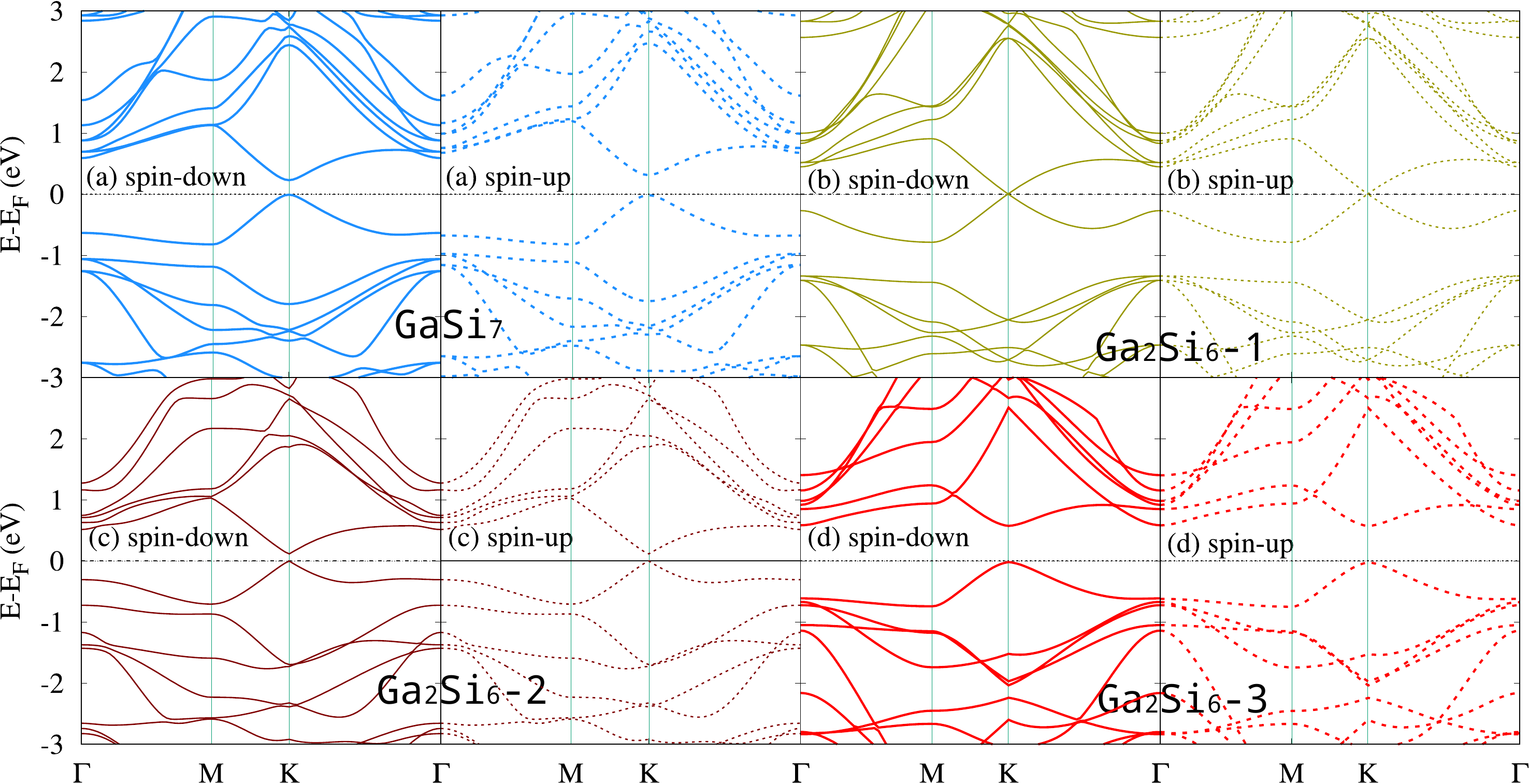}
	\caption{Band structure for optimized structures of GaSi$_7$ (a), Ga$_2$Si$_6$-1 (b), Ga$_2$Si$_6$-2 (c), and  Ga$_2$Si$_6$-3 (d) for both spin-down (dashed lines) and spin-up (dotted lines). The energies are with respect to the Fermi level, and the Fermi energy is set to zero.}
	\label{fig03}
\end{figure*} 
The band gap is found to be at a maximum due to the decreased repulsive force when both of the two Ga atoms are placed at para positions in case of Ga$_2$Si$_6$-3, while it is reduced when they are placed at different positions of para and meta in Ga$_2$Si$_6$-2. However, due to the enhanced repulsive force between Ga atoms at para-ortho, the band gap of silicene is found to be very narrow, equal to $0.2$ meV for the Ga$_2$Si$_6$-1 structure.

\subsection{Magnetic properties} 
 
The pure silicene was found to have no magnetism in the system, indicating that it is a nonmagnetic substance. For the varying doping concentrations of Ga atoms in silicene we compute the magnetic characteristics of the resulting material and identify possible ferromagnetic and antiferromagnetic phases. 

We compute the results of a spin-dependent model for the Ga-doped silicene monolayers. 
First, we perform both ferromagnetic (FM) and antiferromagnetic (AFM) calculations with the  magnetization strength assumed to be $0.5 \, \angstrom/$m. The energy difference between the AFM and the FM states is $\Delta E = {\rm E_{AFM}} - {\rm E_{FM}}$.
The $\Delta E$ of the Ga doped silicene is found to be zero for all structures under investigation here indicating a nonmagnetic structures.
The calculations of the spin-dependent band structures can confirm the magnetic properties of the monolayers. The spin-dependent electronic band structures of GaSi$_7$, Ga$_2$Si$_6$-1, Ga$_2$Si$_6$-2, and Ga$_2$Si$_6$-3 for both spin-up (solid lines) and spin-down (dotted lines) are shown in \fig{fig03}. The spin-up and spin-down of band structures are symmetric, and there is no spin splitting between them at the Fermi energy.
It is found that the materials behave as spin unpolarized semiconductors with no spin splitting between the spin-up and spin-down channels. As a result, these materials are known as nonmagnetic semiconductors. 
 
\begin{figure}[htb]
	\centering
	\includegraphics[width=0.35\textwidth]{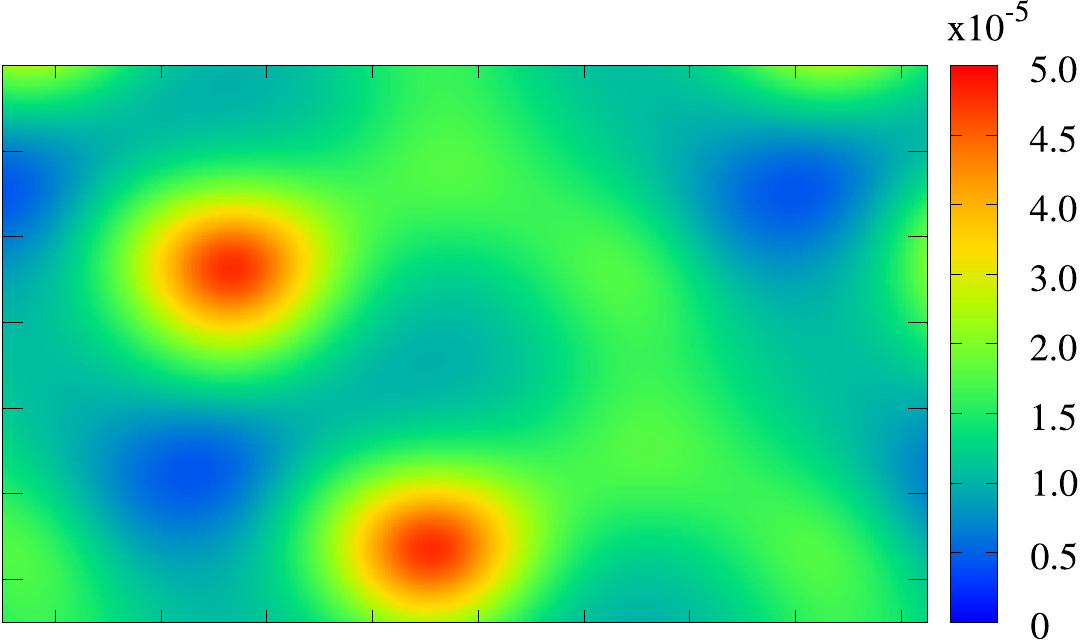}
	\caption{The spin polarized density of Ga$_2$Si$_6$-2 structure.}
	\label{fig04}
\end{figure}

Another confirmation of the non-magnetic properties of these structures is the spin polarized density.
For instance, the spin polarized density of Ga$_2$Si$_6$-2 is shown in \fig{fig04}.
In both situations, FM and AFM, the distance between the atoms remains unchanged, as do the effects of both states on the lattice constant and the distance between the Ga-Ga atoms. 
The spin polarized density of Ga$_2$Si$_6$-2 is extremely small indication a non-magnetic material.

\lipsum[0]
\begin{figure*}[htb]
	\centering
	\includegraphics[width=0.7\textwidth]{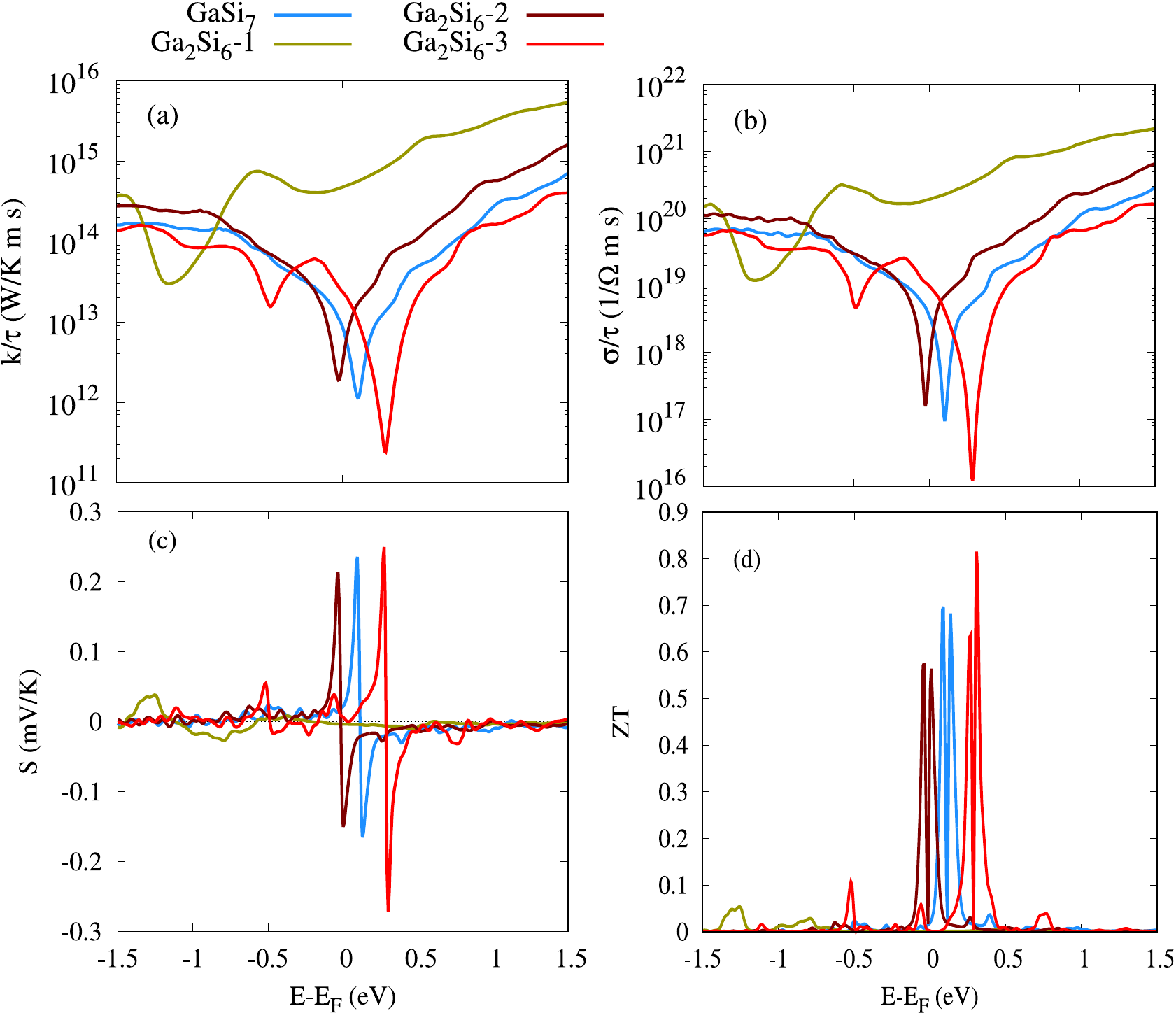}
	\caption{Electronic thermal conductivity, $k$ (a), electrical conductivity, $\sigma$ (b), Seebeck coefficient, $S$ (c), and figure of merit, $ZT$ (d) versus energy are plotted for pure silicene, GaSi$_7$ (blue), Ga$_2$Si$_6$-1 (golden), Ga$_2$Si$_6$-2 (blood), and  Ga$_2$Si$_6$-3 (red) . The Fermi energy is set to zero.}
	\label{fig05}
\end{figure*}

\subsection{Thermal properties} 

Thermal characteristics, such as the electronic thermal conductivity, the electrical conductivity, the Seebeck coefficient, and the figure of merit are examined and shown in this section. The thermal calculations are performed in the temperature range of 20–150 K. The electron contribution to transport is considerable, but the phonon contribution is negligible, and the electron and lattice temperatures are unrelated \cite{Nika_2012, D0NR06824A}.
Scientists have become more interested in the issue of removing heat from nanoelectronic devices and thermoelectrically turning excess heat into energy in recent years. The next generation of nanoscale electronic devices will require this thermoelectric energy conversion \cite{Sadeghi2015}. The figure of merit, which is connected to the Seebeck coefficient, the electrical conductivity, the temperature, and the electronic thermal conductivity, determines the efficiency of heat conversion. It has been proven that a material with a high figure of merit should have a high electrical conductivity and Seebeck coefficient, and low thermal conductivity with a specificity temperature. 
A thermoelectric device's or material's efficiency is evaluated by its thermoelectric figure of merit, which is defined as $ZT = S^2 \sigma T/k$ \cite{nano9020218},
where $S$ is the Seeback coefficient, $\sigma$ is the electrical conductance, $T$ the temperature and $k$ the electronic thermal conductance. A monolayer silicene's thermoelectric performance is low due to closeness of its band gaps. Pure silicene has a low $S$ and a high $k$, resulting in a low \cite{LI2020106712}.
Figure \ref{fig05} indicates the relationship between $k$ (a), $\sigma$ (b), $S$ (c), and $ZT$ (d) with energy. Lowering the $k$, which enhances the structure and increases the $ZT$, is achieved by doping silicene. The decrease in the value of $k$ is found to be significantly higher than the change in $\sigma$. Doping Ga in silicene opens the band gap, which is dependent of the distance between dopant atoms rather than the increase in concentration. The $S$ and $ZT$ in GaSi$_7$, Ga$_2$Si$_6$-1, Ga$_2$Si$_6$-2, and Ga$_2$Si$_6$-3 structures are estimated to rise. The Ga$_2$Si$_6$-1 has a low band gap, but Ga$_2$Si$_6$-3 has the largest band gap, resulting in the highest $S$ and $ZT$. The other structures GaSi$_7$, Ga$_2$Si$_6$-1 fall between them. As a result, the Ga$_2$Si$_6$-3 should perform better in terms of thermoelectricity.

\subsection{Optical properties} 

In this part, we calculate the optical properties, such as the imaginary part of the dielectric function, $\varepsilon_2$, the excitation spectra, $k$, and the optical conductivity for 2D Ga-doped silicene in parallel (E$_{\rm \parallel}$) and perpendicular (E$_{\rm \perp}$) polarized electric fields. The electronic band structure has a direct impact on the optical properties. As a result, changing the band gap of a material has an effect on its optical characteristics.
The dielectric function is widely used to evaluate the optical properties, and it can also be used to describe other qualities. The material's ability to absorb energy is proportional to the imaginary component of the dielectric function. 

There are two primary peaks in $\varepsilon_2$, and $k$ for for silicene with no doping \cite{JOHN2017307}. In the M-K direction, the $\pi-\pi^*$ transition occurs at $1.68$~eV, whereas the $\sigma-\sigma^*$ transition occurs at $3.92$~eV in both the M-K and $\Gamma$-K directions. For the case of E$_{\rm \perp}$, interband transitions have been reported, and the transitions occur at energies below 10 eV \cite{JOHN2017307}.

\begin{figure}[htb]
 	\centering
 	\includegraphics[width=0.4\textwidth]{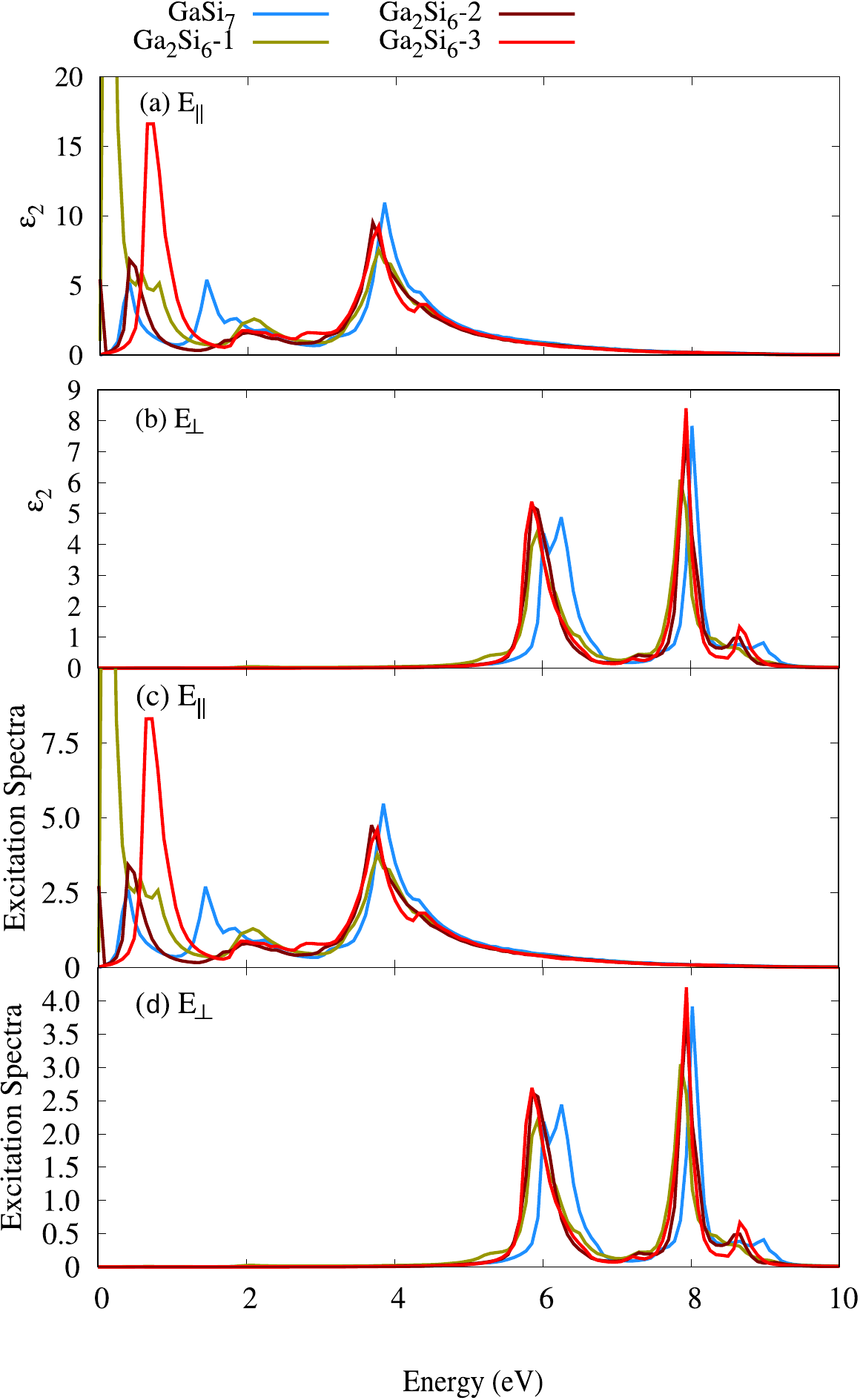}
 	\caption{Imaginary part of dielectric function, $\varepsilon_2$, and excitation spectra, $k$, in the case of electric field that is parallel (E$_{\parallel}$) and perpendicular (E$_{\bot}$) to the GaSi$_7$ and Ga$_2$Si$_6$ structures.}
 	\label{fig06}
 \end{figure}

We show the $\varepsilon_2$ (a,b) and excitation spectra (c,d) for light polarized parallel and perpendicular to the plane of the GaSi$_7$, Ga$_2$Si$_6$-1, Ga$_2$Si$_6$-2, and Ga$_2$Si$_6$-3 in \fig{fig06}. The dielectric function and the excitation spectra are affected by the repulsive interaction of the Ga-Ga atoms as the band structure and the band gap are tuned due to the Ga atoms. Both peaks occur with a little shift to lower energy in both directions, and the high peak is created at low energy for the imaginary part of the dielectric function. Ga$_2$Si6-3 has the highest pick among the structures because of the largest band gap arising from the minimized repulsive force between Ga atoms.

\begin{figure}[htb]
	\centering
	\includegraphics[width=0.45\textwidth]{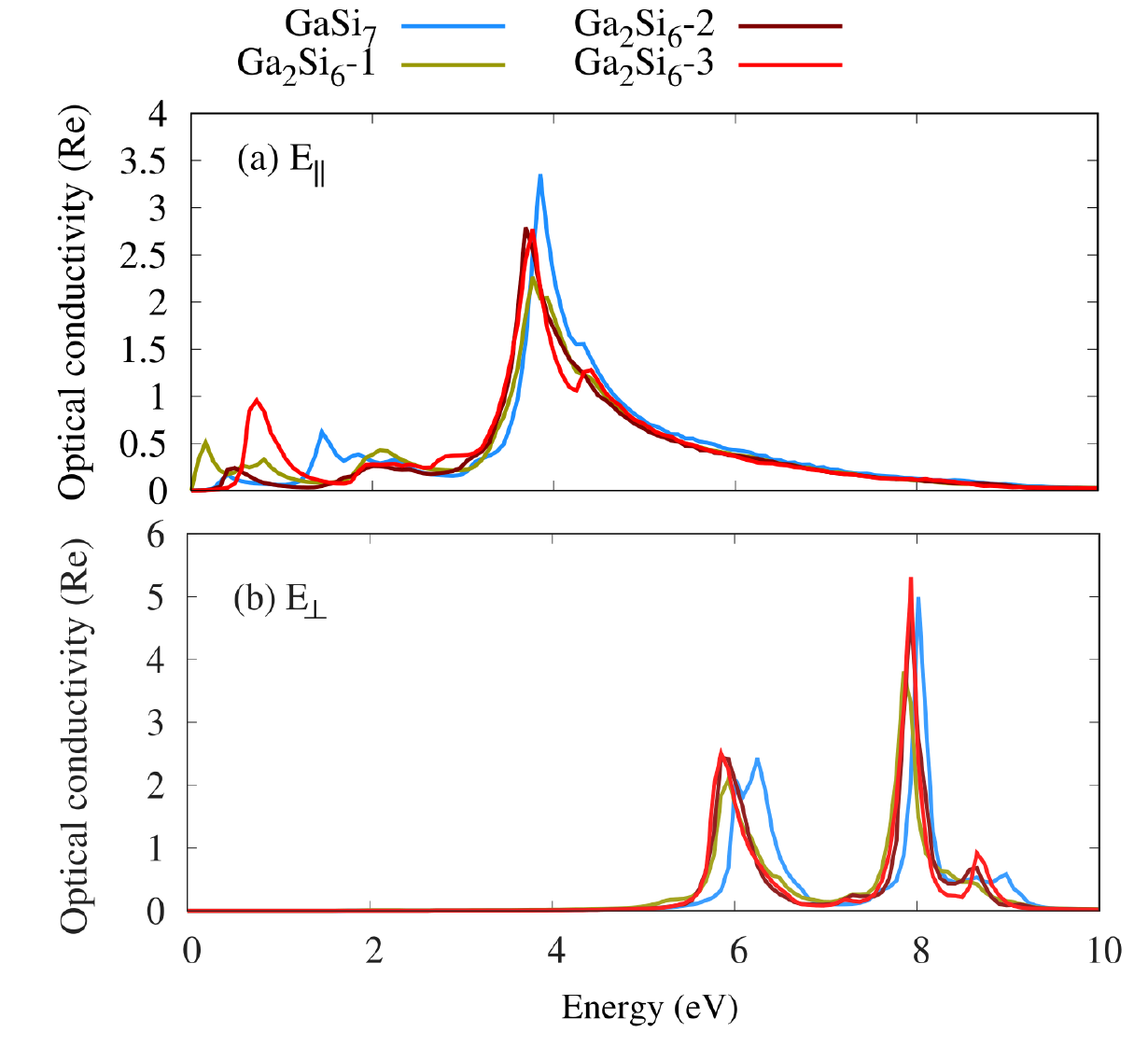}
	\caption{Optical conductivity (Real part) in the case of electric field that is parallel (E$_{\parallel}$) and perpendicular (E$_{\bot}$) to the GaSi$_7$ and Ga$_2$Si$_6$ structures.}
	\label{fig07}
\end{figure}

The last part of this study is the influence of doping on the optical conductivity (real part) of GaSi$_7$, Ga$_2$Si$_6$-1, Ga$_2$Si$_6$-2, and Ga$_2$Si$_6$-3 structures in parallel and perpendicular electric fields as shown in \fig{fig07}. For these structures, the optical conductivity in E$_{\parallel}$ is almost zero up to $0.28$ eV, $0.2$ meV, $0.16$ eV, and $0.6$ eV, indicating the band gap of these monolayers, which have semiconductors properties. The Ga$_2$Si$_6$-3 has the highest peak value in terms of low energy among them due to its repelling action.

\section{Conclusion}

The effects of the atomic configuration, the interatomic distance, and the atomic interaction of the dopant atoms with silicene are given special consideration in this study. Despite its favorable properties, silicene, like graphene, has a zero band gap, which restricts its application in nanodevices. We show that a Ga doped silicene monolayer opens a band gap. The configuration and the dopant concentration ratio control the band gap according to the first-principles calculations. The effects of doping has been examined by changing the dopant concentrations from $12.5\%$ to $25\%$, as well as examining alternative atomic positions for the same replacement doping concentration. The band gap in a 2D silicene doped with Ga is seen to be significantly dependent on the location of the Ga atoms and the Ga-Ga interactions. 
The band gap is determined by the distance between the dopant atoms and the Coulomb repulsive force, rather than the doping concentration. As a result, the band gap in a monolayer of silicene can be widened, resulting in enhanced thermoelectric and optical properties. Furthermore, these structures behave as nonmagnetic semiconductors because they are spin unpolarized and have no spin splitting between spin-up and spin-down levels in the band structure. Finally, our calculations revealed that Ga doped silicene monolayers will be beneficial for thermoelectric and optoelectronic devices. 

%\bibliographystyle{elsarticle-num} 
%\bibliography{Ref_2.bib}

\end{document}